\documentclass[12pt,aps,prd,showpacs,floatfix,nofootinbib,
amsmath,amssymb,
reprint]{revtex4-2}

\usepackage{graphicx}
\usepackage{dcolumn} 
\usepackage{color}
\usepackage{bm}
\usepackage{amsmath}
\usepackage{amsfonts}
\usepackage{hyperref}
\usepackage{natbib}

\usepackage[utf8]{inputenc}
\usepackage[T1]{fontenc}
\usepackage{mathptmx}

\newcommand{\mathbit}[1]{\boldsymbol{#1}}

\begin{document}
	\title[Relativistic Wind Farm Effect: Possibly Turbulent Flow of a Charged, Massless Relativistic Fluid in Graphene]{Relativistic Wind Farm Effect: Possibly Turbulent Flow of a Charged, Massless Relativistic Fluid in Graphene}
	
	\author{Mark Watson}
	
	\affiliation{Department of Physics, University of Colorado, Colorado Springs, Colorado 80904, USA}
	
	\date{\today}
	
	\begin{abstract}	
		At low Reynolds numbers, the wind flow in the wake of a single wind turbine is generally not turbulent.  However, turbines in wind farms affect each other's wakes so that a turbulent flow can arise.  In the present work, an analogue of this effect for the massless charge carrier flow around obstacles in graphene is outlined.  We use a relativistic hydrodynamic simulation to analyze the flow in a sample containing impurities.  Depending on the density of impurities in the sample, we indeed find evidence for potentially turbulent flow and discuss experimental consequences.
		
	\end{abstract}
	
	\maketitle
	
	
	\section{Introduction}

Graphene is a sheet of graphite, one carbon atom thick, with interesting physical and electrical properties.  Its provides an excellent experimental platform to observe electric transport properties and test theories on conductivity at moderately high temperatures.  The two-dimensional solid is made of carbon atoms arranged in a honeycomb lattice, showing a high conductivity through a simple band structure.  Of focused interest is the charge neutral state where each carbon atom contains exactly one electron so that half of the energy levels are occupied.  In this state the band gap disappears at points on the Fermi surface called Dirac points, and the dispersion relation is nearly linear surrounding the points and form a conical Fermi surface at low energies.  Deviations from the linear dispersion begin to appear at a very high temperature, $\approx 10^5 K$, validated experimentally in \cite{Zhang2005}\cite{KrishnaKumar2017}.  The electric charge is transported through the conical band structure by highly mobile, massless chiral quasi-particles \cite{Novoselov2005} referred to as Dirac particles.  The number density of the quasi-particles is determined using electrical properties such as the Hall resistance, which normally assumes an integer value in a two-dimensional solid, but in graphene is found in multiples of $1/2$, known as the fractional Quantum Hall Effect \cite{PhysRevLett.59.1776}.  In the presence of a strong magnetic field the wave functions of the quasi-particles behave like simple quantum harmonic oscillators that take on discrete energy levels based on the strength of the magnetic field.  This effect, called the Shubnikov-De Haas effect, is used to determine the effective mass of the charge carrying quasi-particles.  

The dynamics of the quasi-particles constituting the electric current can be described collectively with hydrodynamics, but since the quasi-particles have no mass and are subject to Dirac’s equation their collective velocity distribution cannot be given by a Maxwell-Boltzmann distribution.  Hence, despite having typical velocities much less than the speed of light, the flow of these particles is better described with relativistic hydrodynamics.  Because of this the study of the electric current in graphene can also be applicable to some astrophysical systems such as gamma ray bursts, supernovae or the flow of particles around the event horizon of a black hole.  These systems share the property of a relativistic hydrodynamic flow, but are of course, much less accessible to experiment.  See \cite{rezzolla2013relativistic} \cite{romatschke2019relativistic} for recent textbooks on relativistic hydrodynamics. 

The transport of the charge carrying quasi-particles within graphene demonstrates a high conductivity that stays above a minimum value, even when carrier concentrations approach zero \cite{Novoselov2005}.  Additionally, the viscosity to entropy ratio is determined to be very small; smaller than that of superfluid helium \cite{PhysRevLett.103.025301}.  A sample of graphene will contain impurities embedded within the honeycomb lattice structure or within the supporting substrate that will affect the current based on their number density, size, placement and electric properties.  In a hydrodynamic flow, if the average fluid speed around these obstructions is large enough compared to the viscous damping effect, a turbulent flow is possible.  

\begin{figure*}[ht] 
	\centering
	\includegraphics[width=2.25in, height=1.85in]{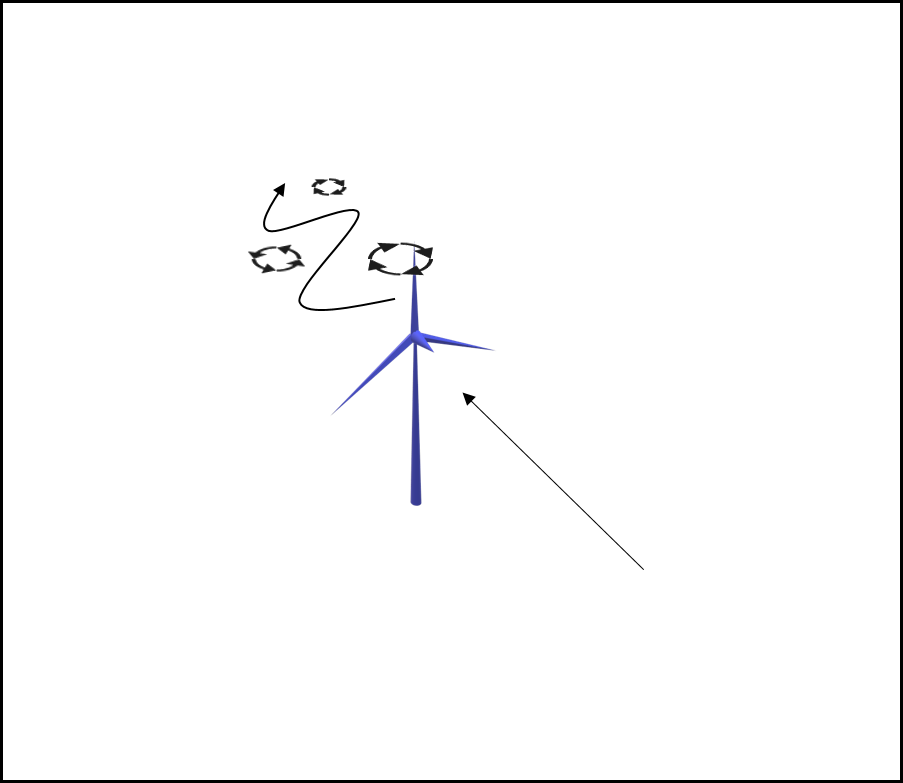}
	\includegraphics[width=2.25in, height=1.85in]{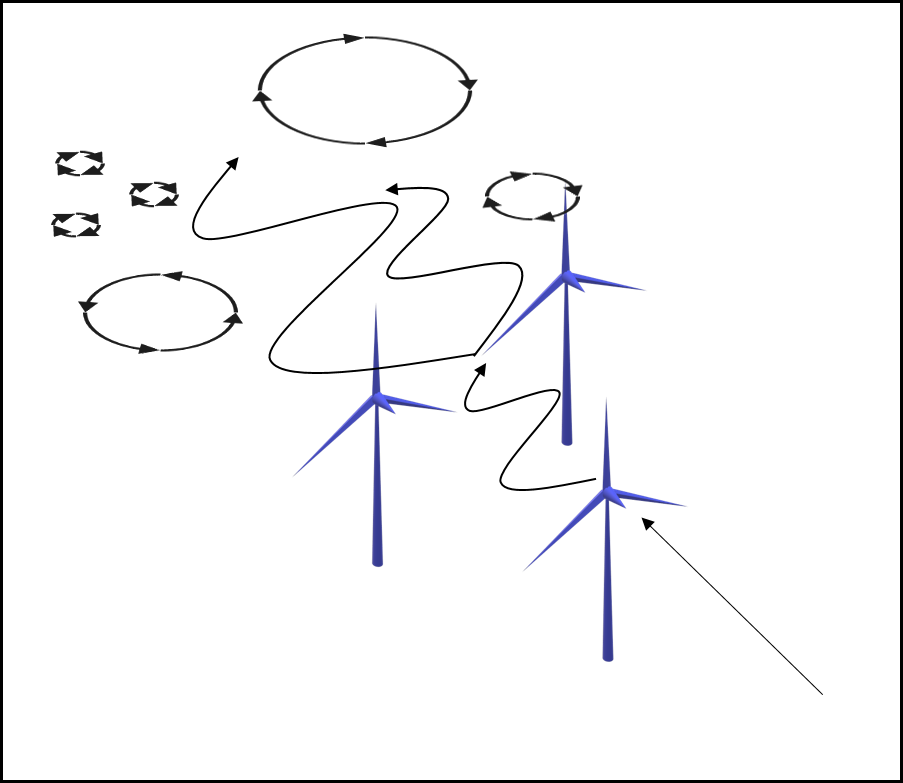}
	\caption{The left shows a vortex street effect a single wind turbine can have on the air flow in its wake when the wind velocity is large enough.  The right sketches a turbulent wind flow caused by additional wind turbines positioned in the wake of another.  The vortex street can turn into a turbulent flow.}
	\label{figure:wind-farm-arrangement}
\end{figure*}

In the present study we note a dependency of turbulence in the Dirac fluid in graphene on the position of an obstruction relative to other impurities present in the sample.  The effect is similar to the susceptibility of a wind farm to turbulence based on the placement of the turbines.  An engineer must take into account the effect the wake a turbine has on downwind turbines when determining their placement.  At high wind velocities a single turbine can create a vortex street, or regularly sized and spaced vortices, within the flow of its wake (fig. \ref{figure:wind-farm-arrangement} left).  If another turbine is placed relatively close to its upstream neighbor near its wake, a turbulent flow is shown to be possible (fig. \ref{figure:wind-farm-arrangement} right) which will affect the efficiency of the farm \cite{stevens2017flow}. 

Turbulence is a non-linear, chaotic flow disruption that results in the creation of vortices that can cause fluctuations in the current.  It is difficult to identify the presence of turbulence qualitatively.  The Reynolds number is the unitless ratio of the characteristic length $L$ and average (macroscopic) fluid speed of the system $U$ to the viscous damping term $\nu$; $Re = \frac{LU}{\nu}$.  It is commonly used to identify the boundary between a turbulent flow and a simple chaotic or Laminar flow.  A large ratio indicates an over-matched damping term, and turbulence is thought to be present when it is larger than $5000$.  If the ratio is smaller than that threshold, but larger than $10$ or $100$, the flow is thought to be in the preturbulent regime.  Any flow with a Reynolds number below that is not believed to be turbulent or preturbulent.  Though charge neutral graphene is an exceptionally conductive material and the charged quasi-particles are able to obtain a large velocity compared to the viscosity, it is believed that it is unlikely to be capable of producing a turbulent flow \cite{Lucas_2018}. 

However, a recent study by Mendoza, Herrmann, and Succi \cite{PhysRevLett.106.156601} modeled a $5\mu m$ impurity obstructing a quasi-particle flow of velocity $10^5$ $m/s$ using a two-dimensional relativistic hydrodynamic numerical model.  They noted a pattern of vortex shedding, a phenomenon of preturbulence, by direct observation and through fluctuations in the current density.  They calculate a Reynolds number on the order of $100$ and proposed that, under these circumstances, graphene could potentially produce a flow in the preturbulent range.  

The present report employs a similar hydrodynamic modeler based on the lattice Boltzmann method to reproduce the flow of a charged Dirac fluid in an experimentally realistic sample of graphene, incorporating impurities embedded within the honeycomb lattice (or in the substrate).  Using experimentally realistic parameters, though the estimate for the Reynolds number in this work is smaller, we find a signal of possible preturbulent flow, corroborating the results of \cite{PhysRevLett.106.156601}.   What is new in this work is that we also consider \emph{multiple} impurities in a single sample and investigate if a relativistic flow around these can give rise to turbulent (as opposed to preturbulent) flow signals.

	\section{Model and Methods}

A turbulent flow is characterized by chaotic changes in pressure and velocity throughout the fluid as a result of large kinetic energy in parts of the fluid that overcome the viscous damping.  The Reynolds number provides the most commonly accepted metric to predict the presence of turbulence.  Classically the Reynolds number is determined, as stated, using the kinematic viscosity $\nu$ defined as the ratio of the dynamic viscosity $\eta$ to the fluid's mass density $\rho$; $\nu = \frac{\eta}{\rho}$.  For fluid systems consisting of massless particles, formulations of the Reynolds number commonly replace the mass density with the entropy density $s$.  Therefore, the expression for the dimensionless viscosity is $\frac{\eta}{s}$, and the Reynolds number is expressed as $Re = \frac{UL}{(c^2 / T) (\eta / s)}$, where $\frac{c^2}{T}$ balances the units.  The Reynolds number found in \cite{PhysRevLett.106.156601} is determined with this formulation.  Alternatively, one is able to retain the classical expression of kinematic viscosity for a fluid of massless particles by defining the mass density in terms of the number density of the quasi-particles and their ``effective'' mass; $\rho = n m_e$.  The number density is determined experimentally using techniques such as the measurement of the Hall resistance, while the effective electron mass is determined through methods such as the Shubnikov-De Haas effect.  The resulting Reynolds number is a ratio incorporating the velocity of a volume of massless particles with respect to inertial mass to the kinematic viscosity defined in terms of the effective mass with respect to the particles' electric properties.  This more classical form of the Reynolds number is more readily compared to that of traditional fluids.  

Turbulence can also be identified through its effect on fluctuations in the current density.  Through non-linear dynamics a turbulent flow creates self-organizing vortices within the fluid, and its effects on the current density are characterized by the emergence of multiple traveling modes in Fourier space.  The non-linear effects in the flow can produce a breaking of the longitudinal symmetry in a Laminar flow causing vortices to spontaneously emerge.  The vortex structures cohere to each other, clockwise rotating structures to other clockwise structures, and counterclockwise to counterclockwise.  The structures eventually detach at larger scales, shedding away from the vortex at periodic intervals.  This is referred to as vortex shedding.  When the vortices begin this pattern of coherence and shedding, the flow creates fluctuations in the current density with a defining spectral signal in frequency space and in wave number space.  When the flow reaches a state of turbulence, the modes migrate from a few initial prominent modes and the spectrum becomes broadband.  If the broadband signal is present in both frequency space and in k-space the flow is considered turbulent as well as chaotic \cite{HoeferPersonalComm}.  However, if the signal is narrow banded in either space, that is, the modes do not migrate temporally or spatially, then the flow (though still potentially chaotic) is not considered turbulent.  Note that while the k-space spectrum is easily determined in a modeled system where the data is directly sampled, it is more difficult to assess the wavenumber of the current density on a real sample.

To explore the possibility of a turbulent flow of the charged quasi-particles in a sample of graphene under realistic conditions we simulate the hydrodynamic equations of motion using an adaptation of the relativistic lattice Boltzmann method described by Romatschke, Mendoza and Succi \cite{2011PhRvC..84c4903R}.  The Relativistic Lattice Boltzmann Model (RLBM) is a hydrodynamic numerical modeler based on kinetic theory \cite{succi2001lattice} \cite{benzi1992lattice}.  It is a variation of the popular Lattice Boltzmann Method (LBM) which is derived from a discretized form of the Boltzmann equation that describes the time evolution of the number density of a group of particles as a probability distribution function $f(x, v, t)$.  The function represents the probability that a given particle is in a particular state in phase space, and the equation is an expression of the conservation of particle number, momentum and energy.  The collision term employed in a lattice Boltzmann model is a greatly simplified version of the collision term defined in Boltzmann's equation using the probability distribution function's relaxation to equilibrium.  In the RLBM, the probability distribution function for a fluid in local equilibrium follows a J{\"u}ttner distribution instead of a Maxwell-Boltzmann distribution.  

In this work the RLBM model reproduces a two-dimensional quasi-particle flow in a $40\mu m $ by $10\mu m$ sample of graphene with one or two rigid impurities obstructing the flow.  The sample is simulated with a $1024 \times 256$ node lattice with a spacing of $0.038 \mu m$.  Initial tests incorporate a single circular impurity with a diameter $D$ of $0.5 \mu m$ placed within the sample in a region near the inflow boundary, referred to as the ``obstruction region''.  Subsequent tests are conducted with a second impurity placed at controlled distances from the first within the same region.  At normal temperatures the impurities within a graphene sample are believed to be charged, creating charge puddles in the surrounding region that affect the electric flow.  The impurities are largely sourced from the substrate \cite{PhysRevLett.98.076602} \cite{Adam18392}, but can also be embedded within the sample itself.  The size of the impurities can vary greatly depending on the foreign material, but they typically stay below approximately $0.5 \mu m$ \cite{Lucas_2018}.  The size and placement of the impurities are difficult to control in an experimental setting, but the model seeks to simulate the effects of one or two quasi-isolated impurities on the current density in ideal but realistic conditions in order to determine if the detection of a turbulent signal is possible.  Therefore the diameter of the obstacle is chosen to maximize its turbulence producing potential while maintaining a realistic size.  The velocity of the charged flow in the Dirac liquid can be relatively high owing to a large effective electric coupling constant.  Flow speeds on the order of $10^2$ $m/s$ are common \cite{10.1038/nnano.2008.268} \cite{doi:10.1063/1.3483130}, but the flow can approach velocities as large as 10\% of the Fermi velocity, $v_F\approx 1.1 \times 10^6 m/s$ \cite{Lucas_2018}.  In order to maximize the possibility of a turbulent signal, the model introduces the largest realizable fluid velocity of $10^5$ $m/s$ ($0.1$ $v_F$) into the sample at the same magnitude along the inflow border.  The borders that are perpendicular to the inflow border use periodic boundary conditions, effectively simulating an infinitely wide sample with multiple, regularly placed obstacles, but at a distance where the wakes created by the obstacles cannot affect each other.  Each lattice node in a region occupied by an impurity implements bounce-back boundary conditions.

The Reynolds number is determined for each test to predict the presence of turbulence using the kinematic viscosity based on the number density as described.  The kinematic viscosity in graphene is found to be $0.000132$ $\frac{\hbar c^2}{eV}$ in natural units ($c = k_B = \hbar = 1$) \cite{PhysRevLett.103.025301}, readily obtained from the value of the diffusion constant for momentum $\eta = 2.633\times 10^3 \frac{eV^2}{c^2 \hbar}$ found in \cite{PhysRevLett.103.025301}, the number density $n = 38.93 \frac{eV^2}{\hbar^2 c^2}$ determined in \cite{PhysRevLett.99.226803}, and the effective mass of a charged Dirac quasi-particle participating in the electric flow, usually given in terms of the mass of an electron $m_e$, where $m_e = 5.11 \times 10^5 \frac{eV}{c^2}$.  The effective electron mass is taken here to be $1.0$ $m_e$.  The diameter of the embedded impurity is the most appropriate choice for the system's characteristic length for the Reynolds number formulation; $U = D$.  

The state data for each node in the lattice is collected throughout the simulation and the local macroscopic moments are determined and recorded.  The current density $\mathbit{j}$ is calculated along the lattice nodes at the outflow border, and the frequency of the fluctuations is determined in Fourier space against time.  The spatial fluctuation of the current density is determined along the lattice nodes in the region of the sample down stream from the current inflow referred to as the ``current density sampling region''.  The spectrum of the current density fluctuations in frequency space and in k-space are recorded and plotted for qualitative inspection to look for evidence of mode generation and migration, indicative of a turbulent flow.  

	\section{Results}

		\subsection{Single Impurity}
\begin{figure*}[t]   
	\includegraphics[width=0.95\linewidth]{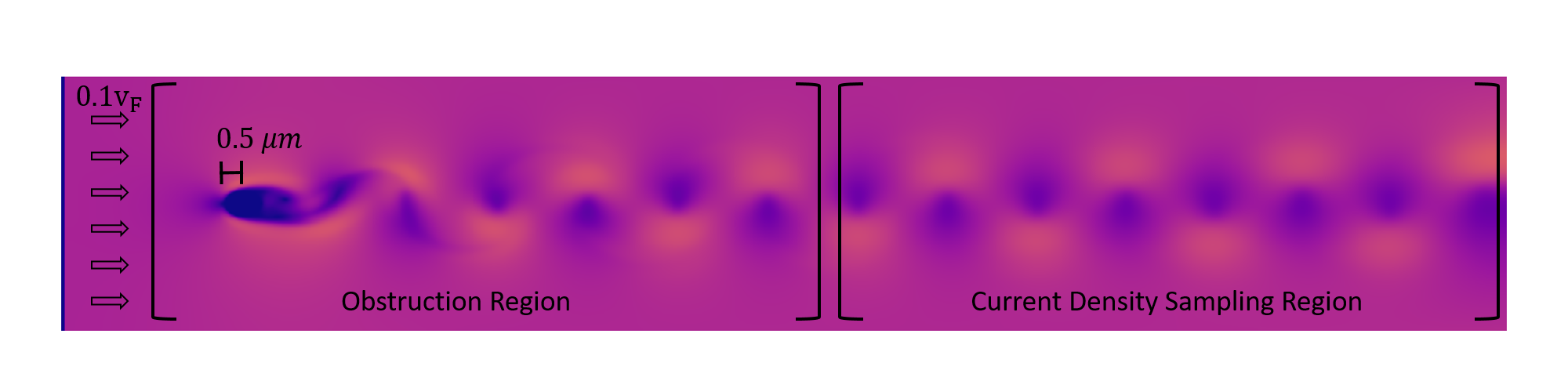}
	\caption{A velocity magnitude heatmap depicting the flow of charge carrying quasi-particles on a simulated sample of graphene at $0.1 v_F$ around an impurity of diameter $0.5 \mu m$ in the obstruction region.  A large velocity around the obstacle creates a regular pattern of vortices in the wake affecting current density fluctuations detected in the sampling region.  The von K{\'a}rm{\'a}n vortex street flow pattern is too regular to be considered turbulent, producing only a single mode in wave number space.}
	\label{single_heat}
\end{figure*}

\begin{figure}[!h]  
	\includegraphics[width=0.8\linewidth]{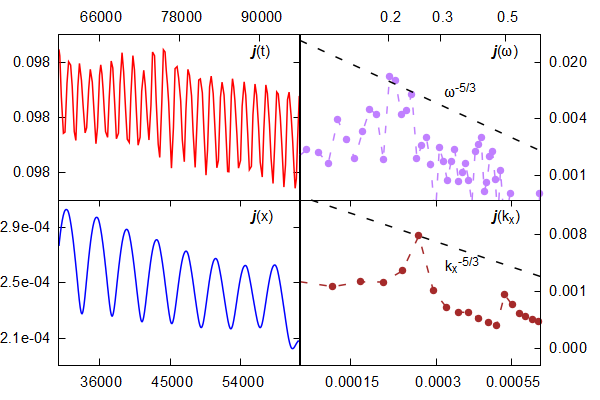}
	\caption{The effects of a single impurity of diameter $0.5 \mu m$ on the current density with respect to time $\mathbit{j}(t)$ (top left) and space $\mathbit{j}(x)$ (bottom left).  The frequency spectrum of $\mathbit{j}(\omega)$ (top right in log scale) shows one or two prominent modes that are beginning to migrate.  One prominent mode is seen in the wave number spectrum of $\mathbit{j}(k_x)$ (bottom right in log scale) indicating the chaotic flow does not show signs of turbulence.  The current density $\mathbit{j}$ is in units of the Fermi velocity $v_F$, and $t, x, \omega,$ and $k_x$ are in modified natural units (see Appendix \ref{appendixA}). }
	\label{single_cur_dens}
\end{figure}

A simulation of a Dirac fluid flow in a sample of graphene containing a single impurity of diameter $0.5 \mu m$ shows a breaking of longitudinal symmetry that develops into vortices in the wake, and forms a flow pattern known as a von K{\'a}rm{\'a}n vortex street.  The vortices form on both sides of the obstacle's wake at a roughly consistent size and placement, alternating on either side of the wake, and shedding at regular intervals (fig. \ref{single_heat}).  The vortex creation, coherence, and shedding produces temporal fluctuations in the current density that are detected at the outflow border and spatial fluctuations found throughout the sampling region (fig. \ref{single_cur_dens}).  The fluctuations produce one or perhaps two prominent modes in the frequency spectrum, but they appear to broaden to create a slope of about $\omega^{-5/3}$ in small portions of the spectrum. There is, however, a single prominent mode in the wave number spectrum indicating the flow around the obstacle does not produce a turbulent signal in the current density.  The Reynolds number for this system is readily calculated in terms of modified natural units based on the Fermi velocity (see Appendix \ref{appendixA}) using the diameter of the impurity as the characteristic length ($L = D = 0.5\mu m = 3562.5 \frac{\hbar v_F}{eV}$), the average flow speed ($U=0.1 v_F$), and the number-density-dependent kinematic viscosity ($\nu = 11.8 \frac{\hbar v_F^2}{eV}$).  It is found to be 
\begin{equation} \label{reynolds_number_ndkv}
	Re = \frac{ L U }{ \frac{\eta}{\rho} } =  \frac{ \left( 3562.5 \frac{\hbar v_F}{eV} \right) \left( 0.1 v_F \right) }{ \left(11.8 \frac{\hbar v_F^2}{eV} \right) } = 6.37.
\end{equation} \\

\subsection{Multiple Impurities}

\begin{figure*}[ht] 
	\includegraphics[width=0.95\linewidth]{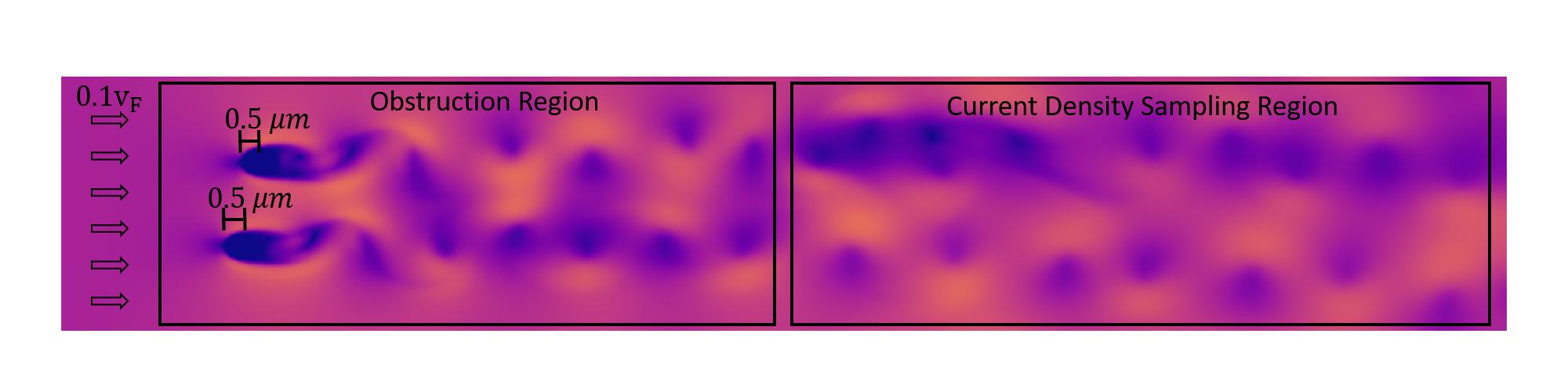}
	\caption{A velocity magnitude heatmap of a quasi-particle flow around two impurities of size $0.5 \mu m$, showing an irregular pattern of vortices in the wake of the two obstacles suggesting turbulence may be present in the flow.  Vortex creation and shedding is evident as in the same system modeling a single impurity, but the irregularity of the coherence of the vortex structures cause a much less regular pattern in the fluctuations in the current density.}
	\label{2obsbest_obs_heat} 
\end{figure*}

\begin{figure}[!h]  
	\includegraphics[width=0.8\linewidth]{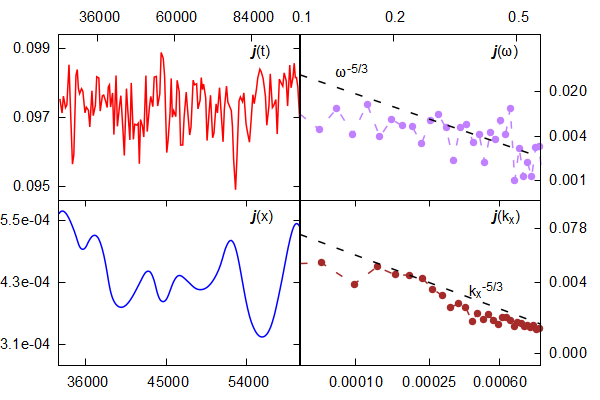}
	\caption{The current density of a quasi-particle flow around two impurities of size $0.5 \mu m$ with respect to time $\mathbit{j}(t)$ (upper left), with respect to space along the axis parallel to the flow $\mathbit{j}(x)$ (lower left), with respect to frequency $\mathbit{j}(\omega)$ (upper right in log scale), and with respect to wave number $\mathbit{j}(k_x)$ (lower right in log scale).  The current density in frequency space, $\mathbit{j}(\omega)$, shows the formation of multiple modes propagating from a single mode creating a rough conformity to a $-5/3$ slope in log scales.  The current density in k-space, $\mathbit{j}(k_x)$, is broadband with a close adherence to a $|k|^{5/3}$ slope expected for migrating modes in a non-linear system.  The broadband spectra in frequency space and in k-space hint at a potential turbulent signal.  The current density $\mathbit{j}$ is in units of the Fermi velocity $v_F$, and $t, x, \omega,$ and $k_x$ are in modified natural units (see Appendix \ref{appendixA}).
	}
	\label{2obsbest_cur_dens}
\end{figure}

A second obstacle is introduced in a subsequent model with the same impurity size and lattice spacing as in the initial test, $0.5 \mu m$ and  $0.038 \mu m$ respectively.  The new obstacle is placed next to, but slightly offset behind the first with respect to the flow in the obstruction region at a separation distance of about $3.38 \mu m$ (fig. \ref{2obsbest_obs_heat}).  The Reynolds number for this system, found to be $12.74$, is similar to the ratio determined for the initial, single-obstacle test, falling to the edge of the lower boundary of what can be considered preturbulent.  However, we see a less regular vortex pattern and the von K{\'a}rm{\'a}n vortex street flow pattern is no longer present.  The fluctuations in the current density show emerging modes in both frequency space and in wave number space which appear to broaden, forming a spectral slope conforming to the power of $-5/3$ as they migrate (fig. \ref{2obsbest_cur_dens}).  There is one or two slightly prominent outlier modes in frequency space, but the higher wave number modes are broadband.  The spectrum in k-space conforms well to a $|k|^{-5/3}$ slope that one might expect for mode creation in a nonlinear system.  Comparing the current density fluctuations in fig. \ref{single_heat} with fig. \ref{2obsbest_obs_heat}, we conclude that the resulting flow in this arrangement of obstacles, despite a borderline disqualifying Reynolds number, may be considered turbulent.

\begin{figure}[h]
		\centering
		\includegraphics[width=0.7\linewidth]{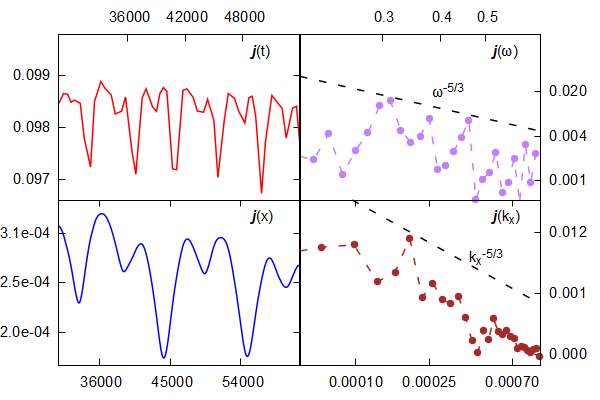}
		\caption{Current density fluctuations recorded from a modeled flow obstructed by two impurities positioned relatively close.  The proximity produces a single dominant mode in wave number space (shown in log scale) similar to what is seen in $\mathbit{j}(k_x)$ from a single obstacle (fig. \ref{single_cur_dens}), implying the effect of closely placed obstacles on the current density is similar to that caused by a single obstacle.  The multi-modal spectrum in frequency space (shown in log scale) is also similar the $\mathbit{j}(\omega)$ spectrum for a single obstacle.  The units of the plot are indicated in Appendix \ref{appendixA}.}
		\label{2obs2close_dens}
	\end{figure}

\begin{figure}[h]
		\centering 
		\includegraphics[width=0.7\linewidth]{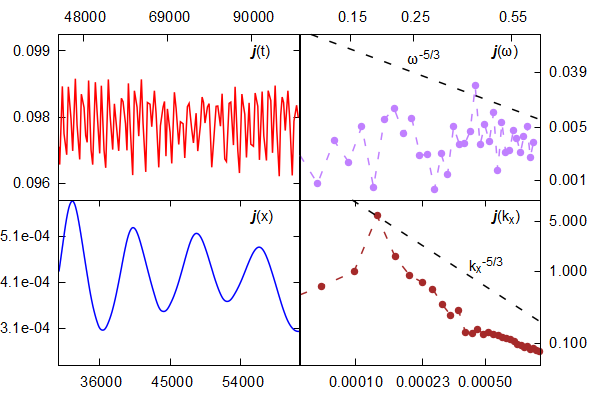}
		\caption{Current density fluctuations from a model simulating a flow around two impurities situated at a large distance relative to their diameter.  In this arrangement two discrete modes appear prominent in the wave number spectrum of $\mathbit{j}$ shown in log scale, and multiple modes are visible in the frequency spectrum of $\mathbit{j}$, also in log scale.  A turbulent signal is not detected for obstacles placed too far apart.  The units of the plot are indicated in Appendix \ref{appendixA}.}
		\label{2obs2far_dens}
	\end{figure}

We find the presence of turbulence in a two-dimensional solid such as graphene to be sensitive to obstacle placement.  A series of subsequent tests place the second obstacle at various other distances from the first in directions both parallel and perpendicular with respect to the inflow.  The second obstacle is positioned such that the flow on the front side of the obstacle is impacted to some degree by the wake created by the first.  When the trailing obstacle is positioned within approximately $2 \mu m$ of the leading obstacle no turbulent or preturbulent signal is detected in $\mathbit{j}$.  At this range the current density's wave number spectrum shows a single dominant mode, and the frequency spectrum shows a few distinct prominent modes with initial signs of broadening, but is not broadband (fig. \ref{2obs2close_dens}).  The spectra are very similar to that created by a single obstacle (see fig. \ref{single_cur_dens}), implying the close proximity of the obstacles has the same effect on the current density as a single, larger obstacle.  Additionally, there is no evidence of a turbulent signal when the obstacles are situated at a large distance with respect to the characteristic length of the system.  For models examined in this work, a turbulent signal is not detected when the obstacles are separated at distances greater than $5 \mu m$.  The current density fluctuations recorded in a simulation of a model with obstacles $6.02 \mu m$ apart show a single prominent mode in wave number space but a somewhat broadband spectrum in the high frequency range in frequency space, with the exception of a single outlier mode; also similar to the current density fluctuations caused by a single obstacle (fig. \ref{2obs2far_dens}).  The contribution of the non-linear effects caused by the wake of first obstacle evidently dissipates at larger distances.


\begin{figure}[h] 
	\includegraphics[width=0.8\linewidth]{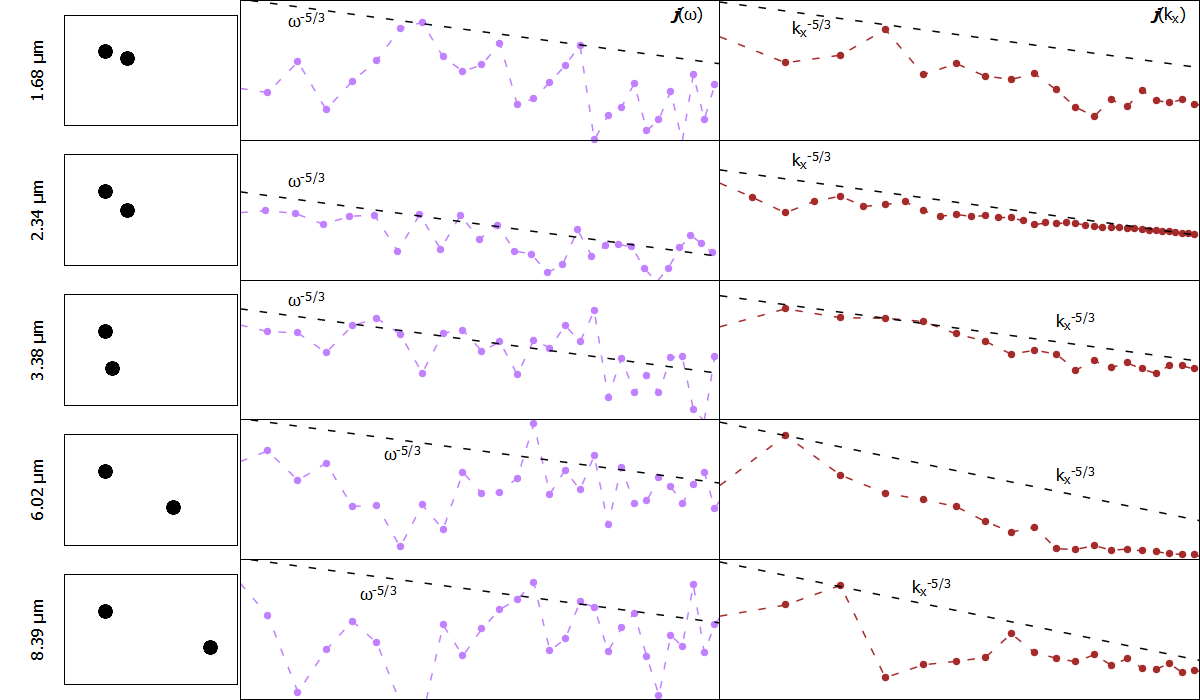}
	\caption{The effect of object separation of two impurities on the current density spectra.  A turbulent signal is detected when the spectrum in both frequency space ($\mathbit{j}(\omega)$ in log scale in the second column) and wave number space ($\mathbit{j}(k_x)$ in the third column; log scale) are broadband.  There is no turbulent signal when the impurities are less than approximately $2 \mu m$ apart as evidenced by the lack of a broadband signal in frequency space.  A broadband signal is similarly absent when the obstacles are separated by greater than approximately $5 \mu m$. }
	\label{2obs_many_cur_dens}
\end{figure}

Fig. \ref{2obs_many_cur_dens} depicts the current density in frequency space and in k-space of five runs with increasing obstacle placement.  At the smallest distance, $1.68 \mu m$, the frequency space spectrum is multi-modal while the k-space spectrum has a single prominent mode, resembling the spectral effects of a single obstacle.  At a slightly larger distance, $2.34 \mu m$, the k-space spectrum becomes broadband and conforms to the slope of  $|k|^{-5/3}$.  More modes are present in the frequency spectrum, but it is not definitively broadband.  A broadband signal is present in both frequency space and wave number space for objects placed at a distance of $3.38 \mu m$ (also shown in fig. \ref{2obsbest_obs_heat}), indicating turbulence.  As the distance increases from $3.38 \mu m$ to $6.02 \mu m$, the frequency spectrum reverts back to multiple modes, and the broadband spectrum in k-space dissipates into a single mode.  The turbulent signal is gone when the objects are separated at this distance.  At the largest separation tested, $8.39 \mu m$, a small number of prominent modes are present in both spectra of $\mathbit{j}$ so that it also resembles the current density spectra created by a single obstacle.  Further tests investigating different placement configurations (not shown) indicate that detection of turbulence is sensitive to other positional features such as alignment.  Because one obstacle must be within the influence of the other's wake, if the objects are adjacent and too far apart, a turbulent signal is not detected.

Note that the identification of a broadband spectrum contains an element of judgment, and a determination for the presence of turbulence using the spectrum is not rigorous.  Indeed, \cite{PhysRevLett.106.156601} indicates the presence of vortex shedding is evidence of preturbulence since it is characteristic of preturbulent phenomena.  Because each model configuration shows multiple modes in frequency space, and because the most prominent mode for each in k-space could be considered an outlier, it is possible that each depicted obstacle configuration can be considered capable of producing a turbulent signal.  Increasing the number of lattice nodes in the model will produce a higher resolution spectrum and subsequently more confidence in the interpretation, but it would also be more computationally expensive.

	\section{Conclusions}


To summarize, we were able to detect a potential turbulent flow in the relativistic hydrodynamics of massless charged quasi-particles in an ultra pure, idealized sample of graphene.  Though the Reynolds number of the modeled systems were on the lower edge of the preturbulent range, evidence of turbulence emerges as a result of interactions in the wake of multiple obstacles placed in the sample.  For two similarly sized impurities a signal of turbulence is dependent upon their separation and position, while apparently independent of size or shape.  This dependence creates a more complex formulation of turbulence that cannot be represented by the Reynolds number metric alone, and has consequences for the conductivity of graphene.  The sensitivity to turbulence based on impurity placement is similar to the sensitivity of the placement of wind turbines in a wind farm.  A notable difference, though, is the importance of higher altitudes above the turbines on the air flow through the farm.  The electric flow within two-dimensional graphene is not affected by out of plane effects.  

A preturbulent signal was detected in a test modeled after \cite{PhysRevLett.106.156601}, in agreement with the findings of that work.  Further investigation is needed to study the effect of semi-rigid obstacles and momentum relaxation on the production of a turbulent signal.  The resonance of a semi-rigid obstacle caused by the flow would amplify the current density fluctuations, and the effect would likely be multiplied by the presence of additional semi-rigid obstacles.  However, the effect would be balanced or mitigated by momentum relaxation effects caused by the obstacle and the lattice itself.  The lattice structure of another two-dimensional solid such as a Kagome metal should also be investigated with the employed RLBM numerical modeler.  The two-dimensional molecular lattice of a Kagome metal creates a linear dispersion relation at low energies, similar to that of graphene, but it has a much stronger electric coupling constant that enables a faster Dirac fluid flow \cite{2019arXiv191106810D_ORIG}.  With a similar viscosity, this implies a higher likelihood of detecting a preturbulent or perhaps a turbulent flow. 

We thank Paul Romatschke for his instruction, helpful discussions and expertise on RLBM modeling.  We would also like to thank Andrew Lucas and Mark Hoefer for their helpful comments and insight.  This material is based upon work supported by the U.S. Department of Energy, Office of Science, Office of Nuclear Physics program under Award Number DE-SC-0017905.  All simulations were run on the Eridanus cluster at the University of Colorado Boulder.  

\section{Data Availability}
The data that support the findings of this study are available from the corresponding author upon reasonable request.

\appendix
	
	\section{Model Scaling} \label{appendixA}

As a technical note, the selected scaling of the RLBM numerical scheme employed in this study is with respect to the Fermi velocity, so the numerical parameters are given in a slightly modified form of natural units where $v_F = k_B = \hbar = 1$.  The factor $\frac{c_F}{c} = 3.333 \times 10^3$ converts all values in natural units of $c$ (where $c$ is the speed of light) to the modified form of natural units.  Length parameters are given in terms of $ \frac{\hbar v_F}{eV}$ and the current density $\mathbit{j}$ is in terms of Fermi velocity $v_F$.  Units of time, frequency and wave number are $\frac{\hbar}{eV}, \frac{eV}{\hbar},$ and $ \frac{eV}{\hbar v_F}$ respectively. 

	\section{Viscidity} \label{appendixB}

Though the more classical form of the kinematic viscosity using mass density is used to determine the Reynolds number, the viscous term in the employed RLBM model is specified by the ratio of the viscosity over entropy density $\frac{\eta}{s}$, as described, and as is more common for the description of the dissipation of massless particles.  Therefore, the number-density-dependent kinematic viscosity used in the calculation of the Reynolds number is converted to an ``effective'' viscosity-entropy ratio $\frac{\eta}{s} = \frac{T_0}{v_F^2} \frac{\eta}{\rho}$ (where $T_0 = 300K$) for the RLBM model.  This ``effective'' viscosity-entropy ratio is referred to in the numerical modeling parameters as ``viscidity'', denoted by $g$, to identify it as the model's viscous parameter, but differentiating it from a true viscosity-entropy or viscosity-mass density ratio.  The viscidity of the modeled system is determined in \ref{viscidity}.  
\begin{equation} \label{viscidity}
	g = \frac{\eta}{s} = \frac{T_0}{c_F^2} \frac{\eta}{\rho} = \left( 2.5862 \times 10^{-2} \frac{eV}{k_B} \right)\left( 11.88 \frac{\hbar c_F^2}{eV} \right)
	= 0.30724 \frac{\hbar}{k_B} 
\end{equation}

\bibliographystyle{hunsrt}
\bibliography{lambda}
	
\end{document}